\documentstyle[editedvolume, namedreferences, psfig]{gamcrckapb} 

\def\lta{\mathrel{\mathchoice {\vcenter{\offinterlineskip\halign{\hfil
$\displaystyle##$\hfil\cr<\cr\sim\cr}}}
{\vcenter{\offinterlineskip\halign{\hfil$\textstyle##$\hfil\cr
<\cr\sim\cr}}}
{\vcenter{\offinterlineskip\halign{\hfil$\scriptstyle##$\hfil\cr
<\cr\sim\cr}}}
{\vcenter{\offinterlineskip\halign{\hfil$\scriptscriptstyle##$\hfil\cr
<\cr\sim\cr}}}}}

\newdimen\digitwidth
\setbox0=\hbox{\rm0}
\digitwidth=\wd0
\catcode`@=\active
\def@{\kern\digitwidth}

\begin{opening}

\title{GALAXIES WITH DENIS:}

\subtitle{Preliminary star/galaxy separation and first results}

\author{Gary A. Mamon}

\institute{Institut d'Astrophysique de Paris \& DAEC, Obs. de Paris\\
98 bis Bd Arago, F--75014, Paris, FRANCE}

\author{Jean Borsenberger}
\author{M. Tricottet}
\author{V. Banchet}
\institute{Institut d'Astrophysique de Paris}

\end{opening}

\begin{document}

\begin{abstract}
The numerous extragalactic and cosmological
motivations of the DENIS and 2MASS near infrared
surveys are outlined.
The performance of the DENIS survey is estimated from $50\,\rm deg^2$ of high
galactic latitude data ($20^\circ < |b| < 60^\circ$).
Simple star/galaxy separation methods are presented and comparison with
300 visually classified objects as well as COSMOS and APM classifications.
We find that the peak intensity over
isophotal area is an excellent star/galaxy separation algorithm, fairly
robust to variations of the PSF within the frames, achieving 98.5\%
completeness and 92.5\% reliability for $I < 16.5$, in comparison with visual
classification.
A new estimate of the photometric accuracy for galaxies is presented.
The limiting factors for homogeneous galaxy extraction at high galactic
latitudes are completeness and
photometric accuracy in $K$, photometric accuracy in $J$ and star/galaxy
separation in $I$ (also used for classification in $J$ and $K$).
Galaxy counts are presented on $50\,\rm deg^2$. The $I$ counts are in
excellent agreement with a Euclidean extrapolation of the published counts
around $I=16-17$ (more so than in all previous studies), 
and thus point to a high normalization at the bright end,
in contrast with the counts published from the APM and COSMOS plate scans.
The $J$-band differential galaxy counts follow the relation $N(J) = 12 \pm 1
\,{\rm
dex} (0.6\,[J-14])\,\rm deg^{-2} mag^{-1}$.
Extrapolation of these high latitude counts suggest that DENIS will produce
highly 
homogeneous catalogs of $\simeq 6000$ ($K < 11$), $\simeq 700\,000$ ($J <
14.8$) and, $\simeq 1\,000\,000$ ($I < 16.5$) galaxies, respectively with
photometric accuracy of $0.08^m$ in $I$ and $0.20^m$ in $J$ and $K$.
Larger highly 
homogeneous samples are expected with improvements to the camera and 
the algorithms.

\end{abstract}

\section {Introduction}

The DENIS consortium has been imaging the southern sky in the 
$I \,(0.8 \mu{\rm m})$, 
$J \,(1.25 \mu{\rm m})$ and
$K_s \,(2.15 \mu{\rm m})$ wavebands since December 1995.
When the survey is complete, around 2000--2001, we expect to have extracted
tens of thousands of galaxies
in $K$, 
roughly one million
in $J$, 
and 
a few million
in $I$ (see \S\ \ref{discus} below for our estimated sizes of homogeneous,
highly complete, reliable and photometrically accurate galaxy catalogs).

Much of the information in this review has been given elsewhere (Mamon et al.
1997b).
The notable improvements here are improved reliability estimates from 
a much larger visually classified sample, a first-order optimization of
star/galaxy separation yielding a one-half magnitude improvement in the
high completeness/reliability magnitude limit and a more accurate estimate of
the photometric accuracy.

\section{Prospective scientific impact}

Wide-angle near infrared (hereafter NIR) galaxy surveys, such as DENIS and
2MASS (see Schneider, Jarrett, Rosenberg and Cutri, all in these proceedings)
will have a wide array of scientific prospects, of which a few are listed
below.  The two important advantages of NIR selection are 1) the near
transparency of interstellar dust in our foreground Galaxy and within
external galaxies, and 2) the low sensitivity of NIR light to recent star
formation in galaxies (see Mamon et al. 1997b), hence a better estimation of
the stellar mass content of galaxies in the NIR.

\paragraph {Statistics of NIR properties of galaxies:}
DENIS and 2MASS will provide the first very large galaxy databases with NIR
photometry.
Photometry of the brighter galaxies will be coupled with redshift
measurements, either already made, or performed during spectroscopic
followups (see, e.g., Mamon 1996; Paturel, in these proceedings) to be used
for  
distance estimates and computation of precise parameters of the fundamental
plane and Tully-Fisher relations (see Vauglin et al. 1997; Rosenberg, in
these proceedings).

\paragraph {Cross-identification with other wavelengths:}
The extragalactic objects extracted by DENIS and 2MASS will be 
cross-identified with analogous samples at other wavelengths, 
such as optical galaxy samples, for example in the Zone of Avoidance 
(see Kraan-Korteweg et al., in
these proceedings), IRAS galaxies (Saunders et al. 1997), 
quasars (see Cutri, in these
proceedings),
radio-galaxies, galaxies found in blind HI surveys (see Kraan-Korteweg et
al., in
these proceedings), etc.
The NIR properties (mainly their location in color-color diagrams) of such
objects will be targeted for discovering new large samples of such objects.
One should expect followups at non-NIR wavelengths of DENIS and 2MASS
galaxies.

\paragraph{Galaxy counts:} 
There has been a debate on the level of galaxy counts at the bright end, as
first estimates (Heydon-Dumbleton et al. 1989; Maddox et al. 1990) found 
a depletion relative to the extrapolation
of the faint-end counts, while later work (e.g. Bertin \& Dennefeld 1997)
disputed this.
This debate has consequences on galaxy evolution and on
whether the environment of the Local Group is
underdense on very large scales ($z \!\!\!\!\lta\!\!\!\! 0.1$). 

\paragraph {Zone of avoidance} 
There are two main applications for
studying galaxies behind the Galactic Plane   (see
Kraan-Korteweg et al., in these proceedings):
1) Mapping the large-scale distribution of galaxies in this still poorly
known region. Indeed, the Zone of Avoidance contains interesting structures
such as the largest 
large-scale concentration of matter in the local Universe, the Great
Attractor (at the intersection of the Supergalactic Plane and the Galactic
Plane, Kolatt, Dekel \& Lahav 1995) and within the Great Attractor, the Norma
cluster, Abell 
3627, richer 
and closer than the Coma cluster (Kraan-Korteweg et al. 1995).
2) The fluxes and angular sizes of galaxies are affected by extinction from
dust in the Galactic Plane, and one can measure this extinction from galaxy
counts (Burstein \& Heiles 1982), colors (Mamon et al. 1997a), and
color-color diagrams (Schr\"oder et al. 1997, and Kraan-Korteweg et al.,
in these proceedings).

\paragraph {Small-scale structures of galaxies}
Only a few catalogs of clusters (Lumsden et al. 1992; Dalton et al. 1997;
Escalera \& MacGillivray 1995, 1996)
and compact groups (Prandoni, Iovino \&
MacGillivray 1994) are based
upon 
automatically 
selected galaxy samples, which happen to be 
optical and photographic (hence subject to
photometric non-linearities).
Because star formation is probably enhanced by galaxy interactions, one
expects that the statistical properties of pairs, groups and clusters of
galaxies built from NIR selected galaxy catalogs will be different from those
built from optical catalogs.
DENIS and 2MASS will thus have the double advantage of using a NIR
galaxy sampled based upon linear (non-photographic) photometry.
The applications of such NIR-based samples of structures of galaxies are
numerous (e.g. Mamon 1994) and include understanding the dynamics of these
structures, 
their bias to projection effects, their constraints on $\Omega_0$
and the primordial density fluctuation spectrum, their use as distance
indicators, and the environmental
influences on galaxies.

\paragraph {Large-scale structure of the Universe:}
The NIR selection and the linear photometry will also benefit the measurement
of statistics (two-point and higher-order angular correlation functions,
counts in cells, 
topological genus, etc.) of the large-scale distribution of galaxies in the
Universe.
For example, the (3D) primordial density fluctuation spectrum of galaxy
clustering can be obtained from the
two-point angular correlation function (Baugh \& Efstathiou 1993)
 or from the 2D power spectrum (Baugh \& Efstathiou 1994).
Moreover, by the end of DENIS and 2MASS, large-scale cosmological simulations
with gas dynamics incorporated (thanks to which galaxies are properly
identified) will provide adequate 
galaxy statistics in projection that will be compared
with those obtained from the surveys, iterating over the cosmological input
parameters of the simulations.

\section {Galaxy extraction and current galaxy pipeline}

The current galaxy pipeline consists of the following steps:

1) Bias subtraction, flat-fielding, bad pixel mapping and astrometric
calibration (standard DENIS Paris Data Analysis Center
pipeline, Borsenberger 1997);
2) Cosmic ray removal;
3) Extraction of photometric zero-points and airmasses from relevant files;
4) Galaxy extraction using the {\sl \nobreak SExtractor} 
(Bertin \& Arnouts 1996) object extraction software, version
1.2b6a (which includes a neural-network star/\-galaxy separator, Bertin 1996,
whose input parameters are 8 isophotal areas, the maximum intensity and as a
control parameter, the
FWHM of the PSF),
with detection and Kron (1980) photometry
parameters optimized from simulated images.

\section {Star/galaxy separation}

Nevertheless, star/galaxy separation is intrinsically difficult because,
at the galaxy extraction limits $I \simeq 16.5$ (see below), 
DENIS will extract roughly 5.5
times as many stars as galaxies in $I$, at very high galactic latitude ($|b|
\simeq 70^\circ$, see Lidman \& Peterson 1996), and the ratio worsens
considerably at lower galactic latitudes and at brighter magnitudes.

We discuss below the steps towards an efficient star/galaxy separation
method.
For this, we extracted in the $I$ band (which has the best angular
resolution)
classical star/galaxy separation diagnostics such as
isophotal area, peak intensity, and FWHM, as well as the neural-network based
stellarity parameter, in a direct fashion, or using a suitably modified
version of SExtractor that includes a two-dimensional modeling of the PSF
that is used as input to the neural network.

\begin{figure}[ht]
\centerline{\psfig{file=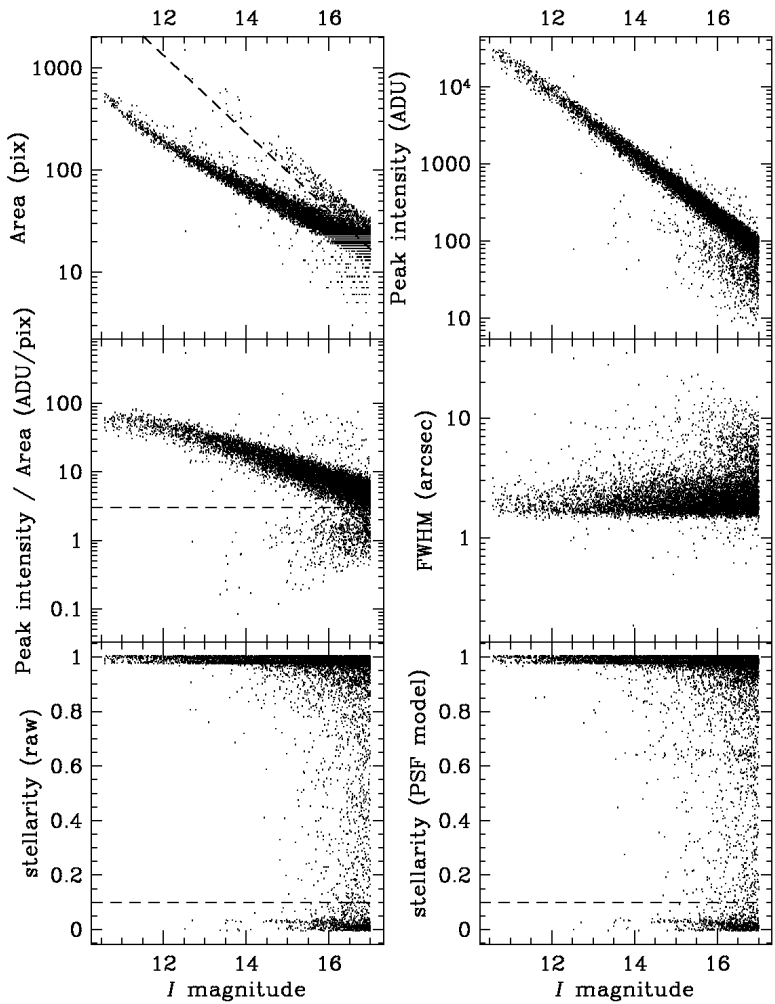,width=0.92\textwidth,angle=0}}
\caption{Diagnostics of star/galaxy separation over one DENIS ($6\,\rm
deg^2$) strip.
The {\it dashed lines\/} are the critical lines for selection of candidates for
visual classification (see \S\ \ref{autosgsep}).
}
\label{diag6}
\end{figure}

Figure \ref{diag6} shows how these quantities vary with magnitude for 
all objects at least 20 pixels from the frame edges on a high latitude strip.

\subsection{Tests of automatic star/galaxy separation}

\label{autosgsep}

One of us (G.A.M.) has classified by eye a set of 329 galaxy candidates on
109 DENIS $I$ band images (of which 33 appeared on consecutive images,
leaving 296 unique candidates).
These candidates were chosen with $I \leq 16.5$, centers at least 20 pixels
from the image borders. Furthermore, they
met {\it at least one\/} of the following loose (to ensure completeness) 
galaxy criteria ({\it dashed lines\/} in Fig. \ref{diag6}):
\begin{itemize}
\item Isophotal area: $A \geq 40 \,{\rm dex} [-0.38  (I-16)]$ pixels
\item Pseudo surface brightness: $\Sigma = I_{\rm peak} / A \leq 3$ ADU/pixels
\item Neural-network stellarity before PSF modeling: $s_0 \leq 0.1$
\item Neural-network stellarity after PSF modeling: $s \leq  0.1$
\end{itemize}

We've used 5 sets of truth tables:
\begin{itemize}
\item Visual DENIS $I$ (see above)
\item COSMOS $b_J$ 
\item APM $b_J$
\item APM $r_F$
\item A mix of the previous 4
\end{itemize}
The COSMOS and APM lists were obtained through the World Wide Web ({\tt
telnet://catalogues@apm3.ast.cam.ac.uk} for the APM and \\
{\tt telnet://cosmos@cosmos.aao.gov.au} for COSMOS).

\begin{figure}[ht]
\centerline{\psfig{file=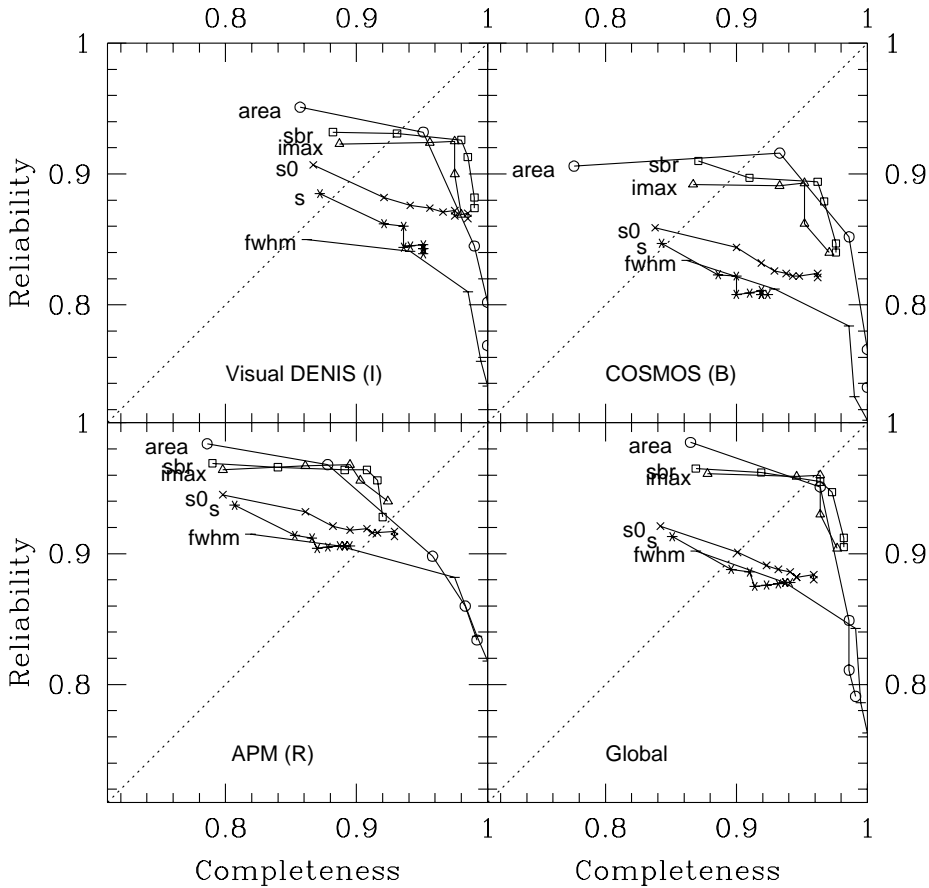,width=\textwidth,angle=0}}
\caption{Completeness versus reliability of different automatic star/galaxy
separation algorithms using 4 different truth tables.
The algorithms are isophotal area ({\tt area}), peak intensity ({\tt imax}),
pseudo-surface brightness ({\tt sbr}), star/galaxy separation without PSF
modeling ({\tt s0}), star/galaxy separation with PSF
modeling ({\tt s}), and full-width half maximum ({\tt fwhm}).}
\label{automatic}
\end{figure}

We've optimized each of the 6 algorithms plotted in Figure \ref{diag6} for a
linear star/galaxy separator in these plots (slope and normalization, except
that we forced a zero slope for the two neural network algorithms).
The results are showed in Figure \ref{automatic}, which plots
the completeness-reliability plots for 4 of
the 5 truth tables.
The different points in Figure \ref{automatic} for a given algorithm
correspond to different cuts through the algorithm versus magnitude diagram
and we only plotted the optimal slope, varying the normalization.

Figure \ref{automatic} shows that the pseudo-surface brightness
criterion is slightly superior to the peak intensity, which, in turn, is
slightly 
superior to the isophotal area (except for the COSMOS-based truth table, for
which isophotal area does best). The other three algorithms (FWHM, and
neural network stellarity before and after PSF modeling), are far inferior
to the first three algorithms.
For the visually classified  DENIS $I$ sample, we achieve 92.5\% reliability
at 98\% completeness, and for the global sample we obtain
96\% reliability at 96\% completeness.
The poor results of the neural networks 
is probably due to the variations of the PSF across the frames, and for
this particular DENIS strip (number 5570), PSF modeling worsened the
results!

\subsection{COSMOS and APM versus visual star/galaxy
separation}

Table 1 shows the comparison between the visual classification
and the classification obtained from the COSMOS
and APM lists.
\begin{table}[htb]
\begin{center}
\caption{Visual DENIS $I$ versus COSMOS and APM star/galaxy separation}
\begin{tabular}{|lr|rrr|rrrr|}
\hline
\multicolumn{2}{|c}{Visual DENIS $I$} & \multicolumn{3}{|c}{COSMOS $b_J$} & 
\multicolumn{4}{|c|}{APM $b_J$} \\
Type & \multicolumn{1}{c|}{Total} & \multicolumn{1}{c}{Galaxy} & 
\multicolumn{1}{c}{Star} & \multicolumn{1}{c|}{Notfound} & 
\multicolumn{1}{c}{Galaxy} & \multicolumn{1}{c}{Star} & 
\multicolumn{1}{c}{Faint} & \multicolumn{1}{c|}{Notfound} \\
\hline
Galaxy	& 203 & 193 & 10 & 0 & 193 & 5 & 3 & 2 \\
Star	& 53  & 6   & 46 & 1 & 9 & 42 & 1 & 1 \\
Star+Star& 8  & 4   & 3  & 1 & 4 & 3 & 0 & 1 \\
Faint	&  21 & 7   & 11 & 3 & 10 & 10 & 0 & 1  \\
Junk    &  11 & 0   & 2  & 9 & 2 & 3 & 3 & 3 \\
\hline
Total  &  296 & 210 & 72 & 14 & 218 & 63 & 7 & 8 \\
\hline
\end{tabular}
\end{center}
\label{apmcosmos}
\end{table}

Of the 11 objects termed as junk, 3 were fragments of a bright galaxy, two
were deemed optical flaws, but according to both APM and COSMOS, one of those
was a star.

The numbers in Table 1 do not permit to establish which
star/galaxy separation is best between visual DENIS, APM or COSMOS.
However, if one assumes that visual DENIS star/galaxy separation is perfect,
one would then conclude that APM and COSMOS both have a completeness of
$193/203 = 95\%$ at $I = 16.5$ (this also assumes that the DENIS $I$
extraction is 100\% complete, which remains to be proven).
The reliability of the extraction would then be
$193/210 = 91\%$ for COSMOS and $193/218
= 89\%$ for APM.

If one assumes that APM or COSMOS are complete, than the incompleteness of
the DENIS galaxy extraction can be estimated from the objects too faint for
DENIS visual classification but called galaxies by the optical surveys.
One obtains completeness levels of 95\% or 97\% at $I = 16.5$ using APM or
COSMOS, respectively.
Of course, if the visual classification were imperfect and that objects
classified as stars or double stars are in fact galaxies, the completeness of
DENIS visual classification would decrease to levels of 90\% or 92\% using
APM or COSMOS, respectively.
Moreover, {\nobreak DENIS} may not have detected objects at $I = 16.5$ that
are seen in 
the optical surveys, and this issue will be addressed in a forthcoming
publication. 

\subsection {Quick and dirty automatic star/galaxy separation}

Since the pseudo-surface brightness criterion seems to produce the best
star/galaxy separation, we have adopted the following preliminary
algorithm for each DENIS
strip:

We adopt a constant critical pseudo surface brightness (independent of
magnitude --- the optimal slope with respect to the visual DENIS $I$, COSMOS
$B$, APM $B$, and global classifications was 0.05), 
by fitting with a cubic polynomial the histogram of the values of
$\Sigma = I_{\rm peak}/A$ for $I \leq 16.5$, in a range chosen to exclude the
peak due to the stars.
Although $k\sigma$ curves down from the stellar locus have negative slope,
the higher slope of the galaxy counts relative to the star counts leads us to 
believe that a given reliability will be achieved with a cut of $k$ that
decreases with magnitude, {\it i.e.\/,} with a lower slope for $\Sigma_{\rm
crit}$. This may explain why the optimal slope is non-negative.

For the $J$ and $K$ bands, we rely on the star/galaxy separation performed in
the $I$ band.
Because the $I$ band has better angular resolution and is more sensitive
than $J$ or $K$ (except at very
low galactic latitudes, corresponding to visual extinction $A_B > 3$, see
Kraan-Korteweg et al., in these proceedings), 
using $I$-band star/galaxy separation 
is superior to doing star/galaxy separation directly in $J$ or in $K$.

Our star/galaxy separation, relying only on  pseudo surface brightness is
simpler than in our previous work (Mamon et al. 1997b), where we required
out galaxies to satisfy both
neural network stellarity (after PSF modeling) and isophotal area
algorithms, and our former star/galaxy separation method
had the disadvantage of using a fixed critical
isophotal area line, whereas strip to strip variations of the PSF lead to
variations of this critical line from one strip to another.

We have thus analyzed a little over $50\,\rm deg^2$ of DENIS data,
restricting ourselves here to $I < 17$.

\section {Photometry}

We estimate below the accuracy of DENIS galaxy photometry using objects
within image overlaps and comparing with APM and COSMOS, and we use
color-magnitude diagnostics as an additional test on the reliability of
star/galaxy separation.

\subsection{Photometric accuracy from overlaps}

\label{overlapsec}

\begin{figure}[ht]
\centerline{\psfig{file=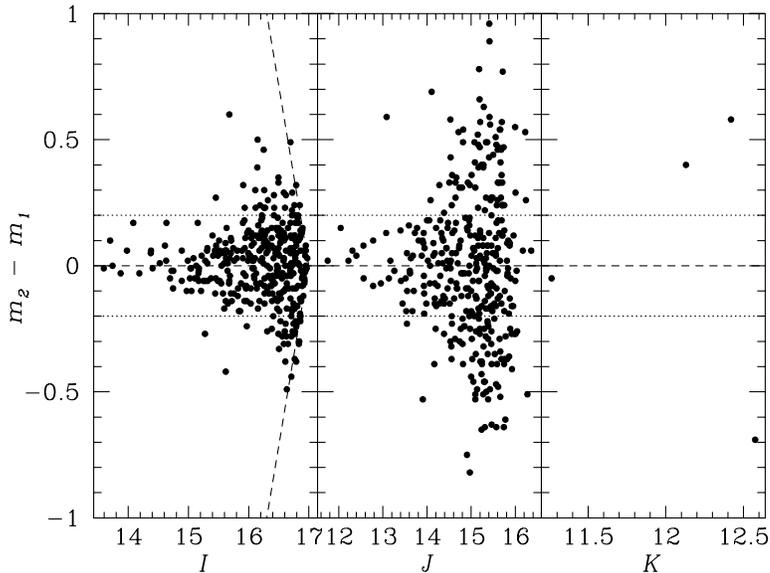,height=8cm,angle=-90}}
\caption{Photometric accuracy for galaxies lying within overlaps of 2 images,
extracted within $50\,\rm
deg^2$ of high galactic latitude ($20^\circ < |b| 60^\circ$) DENIS data.
Objects closer than 20 pixels to the frame
edges are excluded). The  {\it tilted dashed
line\/} represents an $I \leq 17$
selection, whose effects are also seen in the $J$ band.}
\label{overlaps}
\end{figure}

Figure \ref{overlaps} shows the magnitude differences on unflagged
overlap objects
extracted from 50 deg$^2$ of high galactic latitude data.

Contrary to the analogous figure in Mamon et al. (1997b), we have high
certainty on the extragalactic nature of the $J$-band and $K$-band overlap
objects (since again, we rely on $I$-band star/galaxy separation).
For this reason, the photometric accuracy is worse than given in
Mamon et al. (1997b):
The rms error on a single measure is 0.05 at $I = 15$, 0.10 at $I = 17$,
0.10 at $J = 13.7$, and 0.20 at $J = 14.8$.
There are too few $K$ overlaps to
conclude strongly, but indications (based upon only 4 points!)
are that the rms photometric accuracy for
a single measure is roughly 0.20 at  $K \simeq 12.2$.
The $J$-band photometric accuracy was considerably better in our previous
study (Mamon et al. 1997b), but unreliable direct (using neural network
stellarity in $J$ lower than 0.2) star/galaxy
separation had been used for the photometric accuracy study of that work, and
the inclusion of stars  tends to improve the photometric accuracy.

\subsection{Comparison of DENIS galaxy photometry with COSMOS}

\begin{figure}[ht]
\centerline{\psfig{file=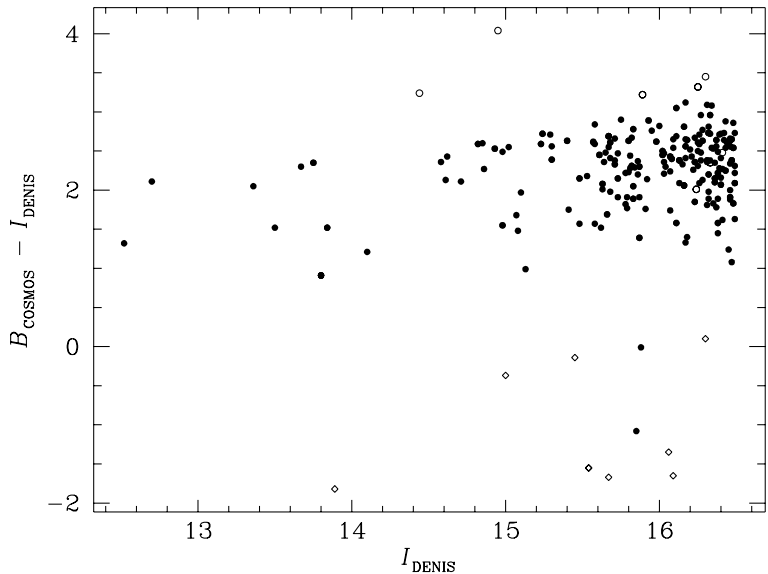,height=8cm,angle=-90}}
\caption{Color-magnitude diagram for galaxies extracted within $3.6\,\rm
deg^2$ of high galactic latitude ($20^\circ < |b| 60^\circ$) DENIS data.
{\it Diamonds\/} refer to objects classified as stars by COSMOS, but as
galaxies by APM, DENIS visual inspection and all DENIS automatic star/galaxy
separation algorithms.
{\it Open circles\/} are objects classified as galaxies with low certainty by
the DENIS visual classification and that were not stars in COSMOS.
}
\label{bmi}
\end{figure}

For the $3.6\,\rm deg^2$ region in which we visually
classified our extracted
objects, we plot in Figure \ref{bmi} the color-magnitude relation obtained
with COSMOS $b_J$ photometry, taken from the World Wide Web.
This figure shows the difficulties in star/galaxy separation, as a number of
points lie far off the $B-I \simeq 2-3$ region.
Part of this difficulty lies in poor star/galaxy separation from COSMOS.
Moreover, there is a trend for bluer galaxy colors at brighter magnitudes,
which we interpret as poor photometry on the {\nobreak COSMOS} side, because
of 
inaccurate compensation for plate saturation.

We also attempted the same with APM data from the World Wide Web, but that
photometry suffers from unusually strong systematic errors at the bright end
(up to 6 mag difference with COSMOS!),
as the photometric calibration has been optimized for stars that saturate at
these magnitudes (Maddox, private communication).

\subsection{Colors of DENIS galaxies}

\begin{figure}[ht]
\centerline{\psfig{file=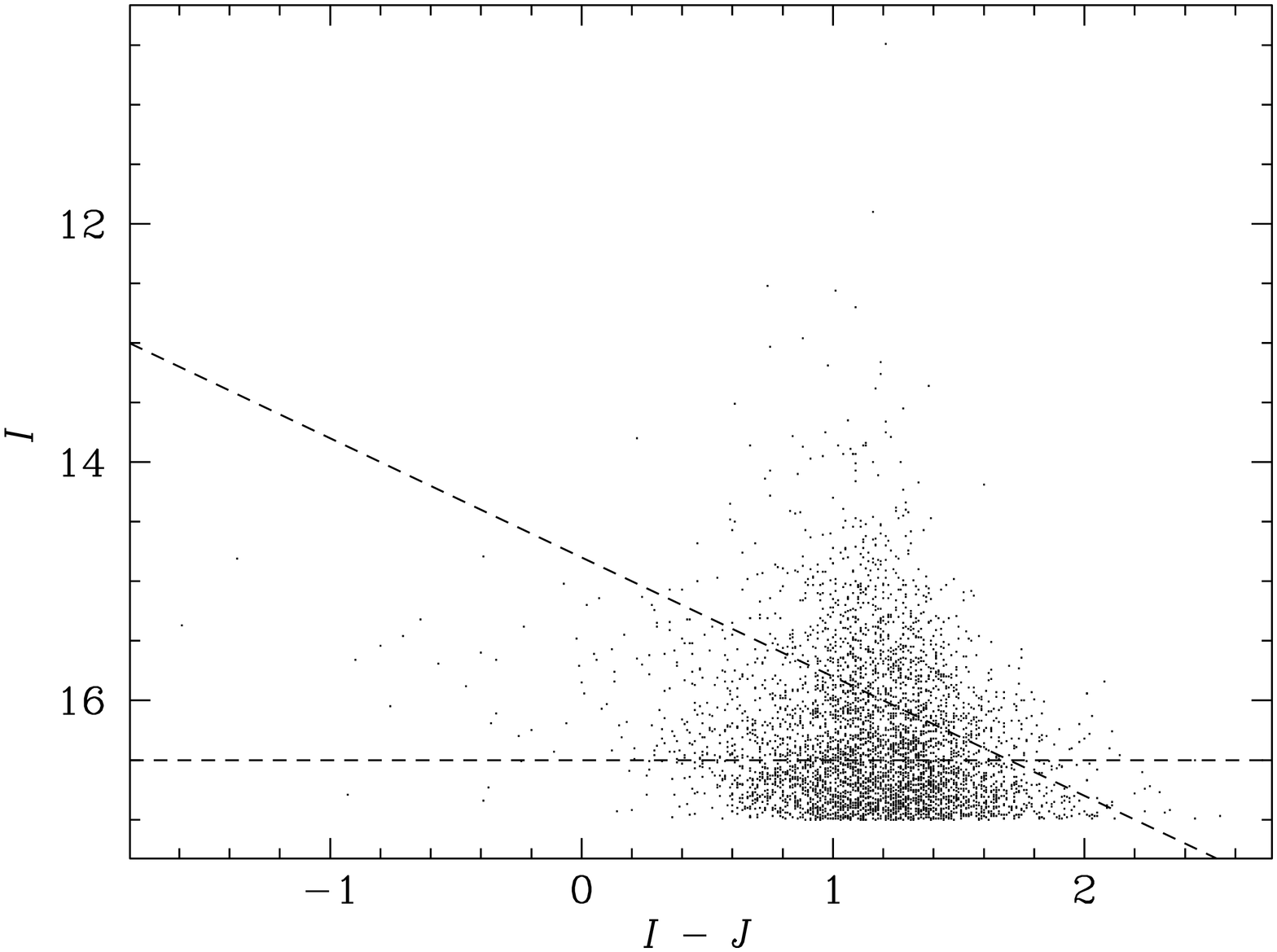,height=9cm,angle=-90}}
\caption{Color-magnitude diagram for galaxies extracted within $50\,\rm
deg^2$ of high galactic latitude ($20^\circ < |b| < 60^\circ$) DENIS data.
Objects closer than 20 pixels to the frame
edges are excluded. The {\it horizontal line\/} represents $I = 16.5$ (the
 and
the dashed line represents $J = 14.8$ 
(the limit for 0.20 mag $J$-band photometry and reliable star/galaxy
separation).}
\label{colmag}
\end{figure}

Figure \ref{colmag} shows the color-magnitude diagram for the galaxies.
The bluest two points turn out to be galaxies!
Visual inspection shows that they are low surface brightness galaxies that
are barely visible in $J$ (and invisible in $K$).
The use of adaptive aperture photometry to define colors makes such objects
appear very blue. We checked that their central colors are normal.

Figure \ref{colmag} shows that at the limit $J = 14.8$
for $\Delta J = 0.20$
mag photometric accuracy, the star/galaxy separation performed in $I$ should
be roughly as reliable as at $I = 16.5$, and could be made even more reliable
by culling out the reddest objects for which $I > 16.5$.

\begin{figure}[ht]
\centerline{\psfig{file=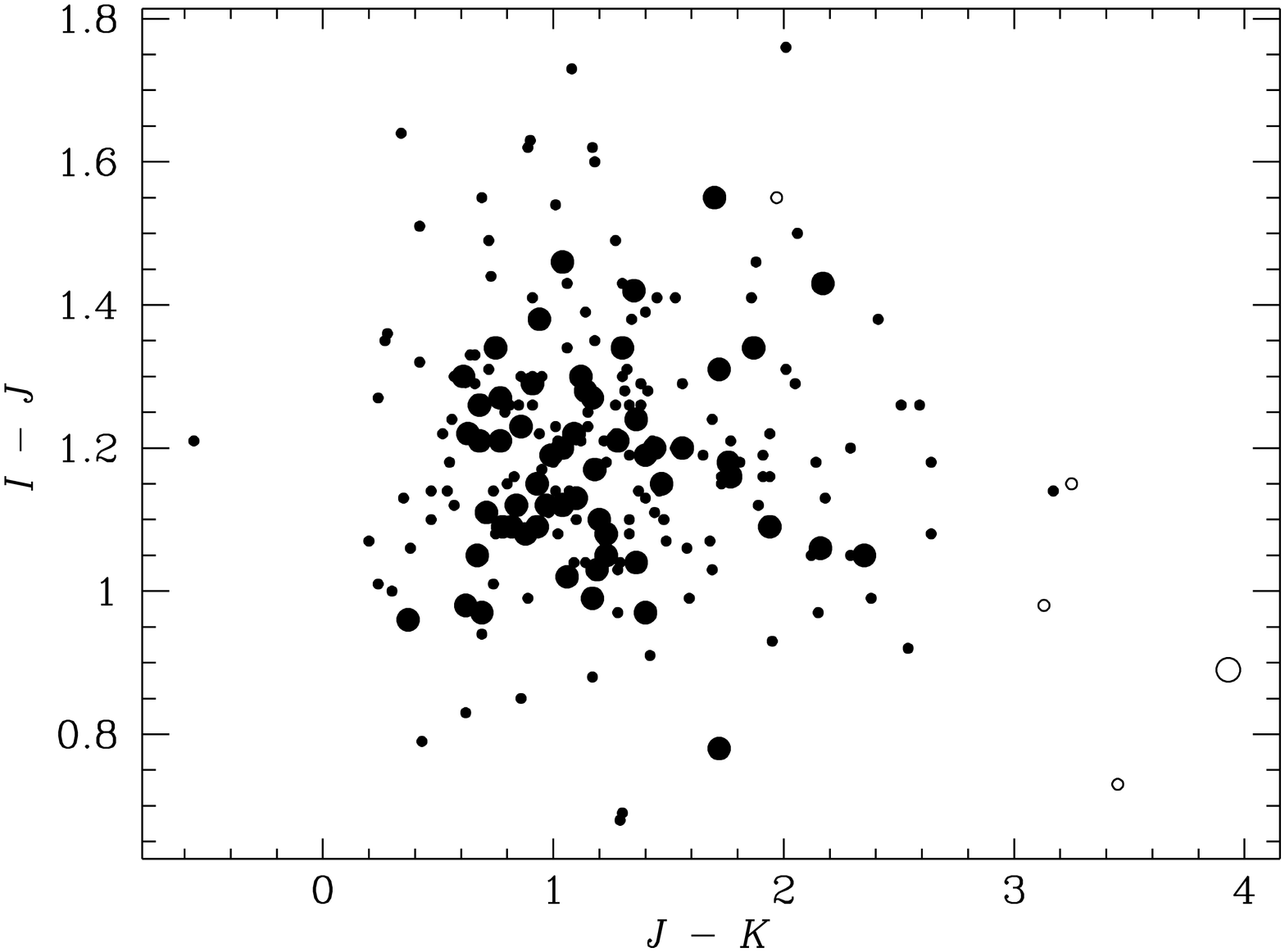,height=9cm,angle=-90}}
\caption{Color-magnitude diagram for galaxies extracted within $50\,\rm
deg^2$ of high galactic latitude ($20^\circ < |b| < 60^\circ$) DENIS data.
Objects closer than 20 pixels to the frame
edges are excluded. {\it Large\/} and {\it small circles\/} are for
objects brighter or fainter than $K = 12$ (the limit for fairly accurate
$K$ photometry, see \S\ \ref{overlapsec}), respectively.
{\it Filled\/} and {\it open circles\/} correspond to objects brighter and
fainter than $I = 16.5$ (the rough limit for reliable star/galaxy separation,
see \S\ \ref{autosgsep}),
respectively. 
}
\label{colcol}
\end{figure}

In Figure \ref{colcol}, we plot the color-color diagram for extracted
galaxies.
The galaxy colors cluster around $I-J = 1.2\pm 0.3$, $J-K = 1.1\pm0.5$, 
but there are indications
for fairly bright objects with red $J-K \simeq 2$ colors, 
which upon visual inspection are
confirmed as galaxies. An important fraction of the points off the
central cluster lie near the frame edges where the PSF is larger.
The large open circle refers to an object too faint in $I$ for reliable
star/galaxy separation, and indeed, visual inspection shows it to be a star
blended with a faint galaxy.

\section{Galaxy counts}

\begin{figure}[ht]
\centerline{\psfig{file=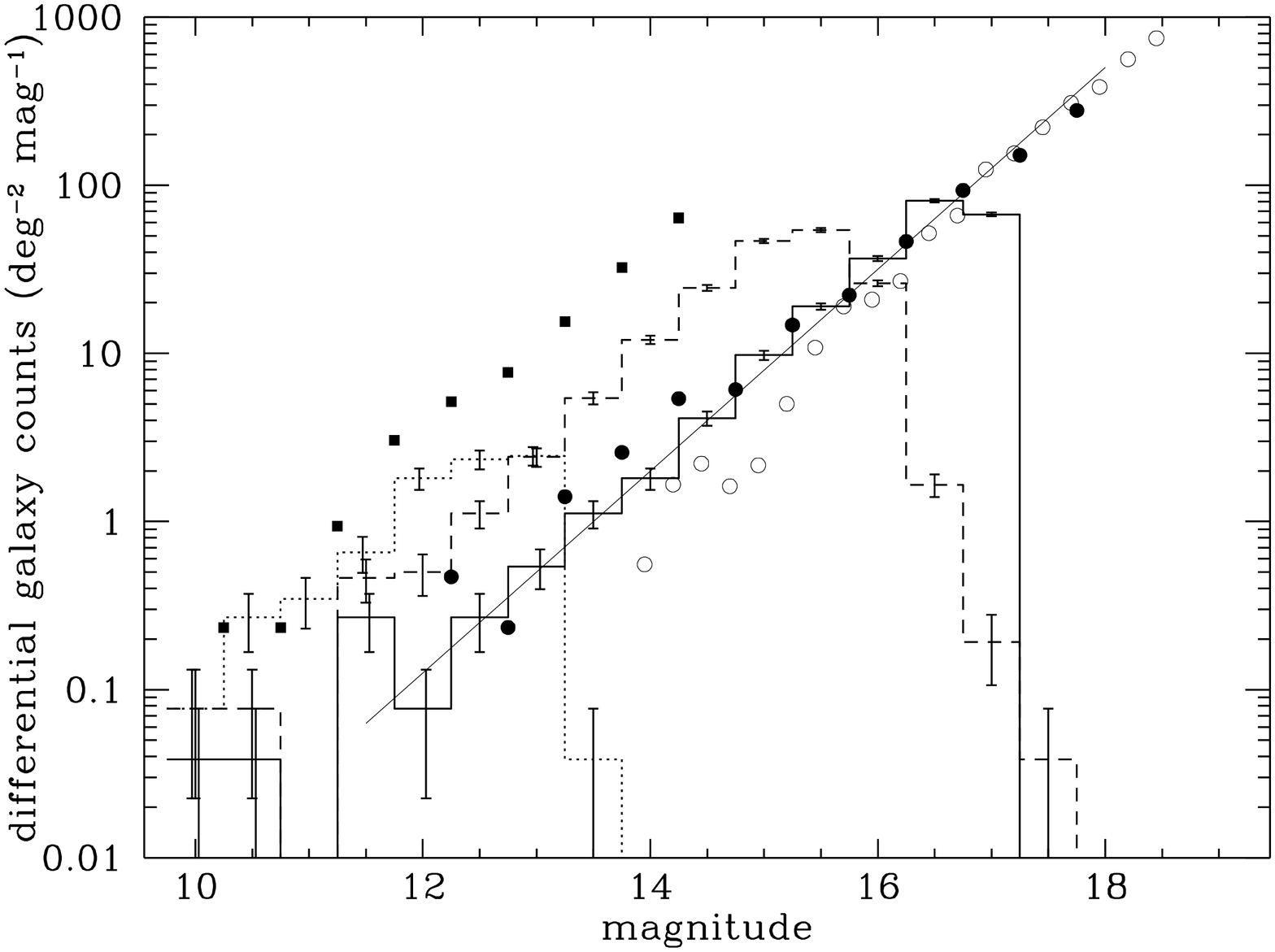,height=10cm,angle=-90}}
\caption{Differential galaxy counts over $50\,\rm
deg^2$ of high galactic latitude ($20^\circ < |b| 60^\circ$), 
$I < 17$ DENIS data.
Objects closer than 20 pixels to the frame
edges are excluded. 
{\it Solid\/}, {\it dashed\/}, and {\it dotted histograms\/} refer to the
DENIS $I$, $J$, and $K$ counts, respectively.
The {\it squares\/} represent $K$-band counts from Gardner et al. (1996),
while the {\it open\/} and {\it filled circles\/} represent the 
$I$-band counts from Lidman and Peterson (1996), and Gardner et al. (1996),
respectively.
The {\it thin oblique line\/} represents an eye-fit Euclidean count function
(0.6 slope).
}
\label{counts}
\end{figure}

Figure \ref{counts} illustrates our $IJK$ galaxy counts.
The $K$-band counts become incomplete at $K \simeq 11$, in comparison to both
published counts 
by Gardner et al. (1996) and to the expected Euclidean 0.6 slope (the
completeness is still roughly 50\% at $K = 12$).

The $I$ band counts match well the published data, although Lidman and
Peterson (1996) find fewer counts at the bright end, while Gardner et al. find
more counts at the bright end (the two sets of published data differ by a
factor of 3 at $I < 15$).
Note that DENIS, Gardner {\it et al.\/} and Lidman \& Peterson all work with
the 
Cousins $I$ band, so no conversion was made from another $I$ filter.
Also, our survey has smaller error bars at the bright end as it covers 
4 to 5 times the solid angle of the two cited
surveys.
Our bright-end $I$-band counts are more consistent with the extrapolation of
the faint counts with a Euclidean slope
than either two sets of published data (our high value at $I=16.5$ is caused
by important stellar contamination in the fainter half of the bin; also, at
$I > 18$, the
published counts become lower than the Euclidean line because of significant
$k$-correction at these magnitudes).
In this sense, although not as high as Gardner et al.'s counts, 
{\it the DENIS $I$-band counts argue for a high bright-end
normalization, consistent with little galaxy evolution at the bright end\/},
in line with analogous findings by Bertin and Dennefeld (1997) using blue
counts. 

The $J$ counts are new (although 
they were already shown in Mamon et al. 1997b).
They are highly complete to $J = 15$, follow very well the Euclidean slope of
0.6, and are well described by the relation
$N(J) \simeq 12 \times {\rm dex}
[0.6 \,(J-14)]\rm\,deg^{-2}\,mag^{-1}$.

\section {Discussion}
\label{discus}

{}From the results of the preceding sections, we can establish limits for the
homogeneous extraction of galaxies from DENIS, as given in Table \ref{limits}.
\begin{table}[ht]
\caption{Estimated DENIS limits from $50\,\rm deg^2$ of reduced data}
\begin{center}
\begin{tabular}{|lccc|}
\hline
	& $I_c$	& $J$ & $K$@ \\
\hline
completeness ($\simeq 80\%$) & 17.25$\,$ & 15.25 & $\,$11@@@ \\
star/galaxy separation (90\% reliability, from $I$) & 16.5@\ & 14.8@ & 13.5@ \\
photometry (0.20 mag accuracy) & $>$18?@@@ & 14.8@ & 12.2?$\,$ \\
photometry (0.10 mag accuracy) & 17.4@ & 13.7@ & $<$11?@@@ \\
\hline
\end{tabular}
\end{center}
\label{limits}
\end{table}

The limiting factors turn out to be 
star/galaxy separation in $I$,
photometry and star/galaxy separation in $J$, 
and
detection in $K$ (assuming that $I$-band star/galaxy separation
is used to classify objects detected in the other bands).

Using the counts from Figure \ref{counts} to extrapolate to the entire survey
area (roughly a hemisphere), we infer that our homogeneous
catalogs will have sizes of $6000$ at $K<11$ (0.2 mag photometry),
$100\,000$ to $500\,000$ at $J < 13.7$ and 14.8
(with 0.2 and 0.1 mag photometry, respectively), and 
$900\,000$ galaxies at $I < 16.5$ (0.1 mag photometry).
The recent installation of an air conditioning system on the $K$ band optics
has decreased the instrumental background by 0.7 magnitude, which should
bring the extraction limit to $K \simeq 11.7$, and thus
increase the size of the homogeneous $K$ 
sample to roughly $15\,000$ galaxies.

Moreover, there is still room for progress on star/galaxy separation.
C. Alard has devised a new algorithm to accurately model the variations of
the PSF across the frame, which need no longer be an elliptical gaussian
(fitting the asymmetric coma of the images), and tests on visually classified
data are about to be performed.
 
\acknowledgements{We thank Emmanuel Bertin for supplying recent updates of
his
SExtractor software package, 
Steve Maddox for useful comments on the  APM  data, Nicolas Epchtein for a
careful reading of the manuscript,
and Pascal Fouqu\'e and the DENIS operations
team.}

\end{document}